

\input amstex.tex
\documentstyle{amsppt}
\magnification\magstep1
\NoBlackBoxes
\topmatter
\title Schwinger terms, gerbes, and operator residues
\endtitle
\author Jouko Mickelsson \endauthor
\affil Department for Theoretical Physics, Royal Institute of
Technology, S-10044 Stockholm, Sweden.
Lectures
at the Stefan Banach Center symposium
"Differential Geometry and Mathematical
Physics", May 15 - 23, 1995, Warsaw. August 31, 1995. \endaffil

\endtopmatter

\define\a{\alpha}
\redefine\e{\epsilon}
\redefine\b{\beta}
\define\g{\gamma}

\redefine\o{\omega}
\redefine\O{\Omega}
\redefine\l{\lambda}
\redefine\L{\Lambda}

\define\gm{\bold g}

\define\<#1,#2>{\langle #1,#2\rangle}
\define\TR{\text{tr}}
\define\dep(#1,#2){\text{det}_{#1}#2}
\define\norm(#1,#2){\parallel #1\parallel_{#2}}

\document
1. INTRODUCTION

\vskip 0.3in
In these notes I want to describe recent developments concerning chiral
anomalies and hamiltonian quantization, their relation to the theory of
gerbes, extensions of generalized loop algebras using the residue calculus
of pseudodifferential operators, and a new renormalization method in QFT.

In $1+1$ space-time dimensions the gauge current is a multiplication operator
acting in the one-particle fermionic Hilbert space $H.$ In particular, the
time components of the current are infinitesimal generators for the group
of gauge transformations. If the physical space is compactified as a circle,
then this group is the loop group $LG.$ A peculiar property of QFT in one
space-dimension is that most interesting operators (including the gauge
current) can be second quantized using a standard normal ordering prescription.
The mathematical reason for this is that the parts of the one-particle
operators mixing positive and negative energy states are Hilbert-Schmidt.

The second quantized operators obey a slightly different algebra than their
classical counterparts. In the process of quantization the Lie algebra $\bold
{gl}_1$ of quantizable operators is promoted to a centrally extended Lie
algebra $\widehat{\bold{gl}_1}.$ The central extension is defined by a
2-cocycle, the Lundberg cocycle [Lu]. On the group level, we have a central
extension of the group $GL_1.$ This includes the loop group $LG$, leading to
a central extension $\widehat{LG},$ the affine Kac-Moody group.

In higher than $1+1$ dimensions new problems arise. The simple normal ordering
method is not sufficient to make the second quantized operators well-defined.
Additional renormalization is needed.  In the case of fermions in background
gauge or gravitational fields the renormalization is achieved by a conjugation
by unitary field dependent operators which reduce the size of the off-diagonal
(with respect to the energy polarization) blocks of the observables.
The size of the operators is best estimated by looking the asymptotic behavior,
in momentum space, of the pseudodifferential symbols. In compact physical
space of dimension $d$ a symbol is of Hilbert-Schmidt type iff it falls to
zero faster than $|p|^{-d/2}$ when $|p|\mapsto\infty.$

After the initial renormalization in the one-particle space the operators can
be
quantized by normal ordering. The Lundberg cocycle leads to an extension of the
original observable algebra. However, now the extension is not simply a central
one because the renormalization is (external) field dependent and therefore
the normal ordering subtraction depends also on the external fields. The
resulting algebra is an extension of the original one by complex functions
of the external fields. At first sight the cocycle is a very complicated
expression in the fields. However, by writing the Lundberg cocycle in a
cohomologically equivalent form, as a residue of an appropriate operator,
we get a simple local expression. The reason is that the residue picks up
only the terms which are of  degree $-d$ in momentum space. The residue formula
for the equivalent of Lundberg cocycle is closely related to the Radul cocycle.
In fact, it can be viewed as a twisted version of the latter, the twist being
defined by the sign of the energy, theorem 5.13. Note that this is only a
partial
cocycle in the Lie algebra of PSDO's. The cocycle property is valid only
for PSDO's $X$ satisfying the Hilbert-Schmidt condition on $[\epsilon, X].$

The plan of these notes is the following.
In section 2 we shall go through some basic definitions in symbol calculus
of PSDO's and the residue of a PSDO is defined and its fundamental properties
are discussed.  In section 3 we discuss the Radul cocycle defining a central
extension of the Lie algebra of PSDO's.  The renormalization needed in higher
dimensional setting is explained in section 4 and is then applied in 5
for computing the structure of the quantized current algebra. In section 6
we take a different, more geometric,  point of view and study the structure of
the bundle of fermionic Fock spaces parametrized by external fields with the
help of the Atiyah-Patodi-Singer index theorem. Finally we explain the
geometry of the bundle in terms of the theory of gerbes. This subsection is
based on recent observations of A. Carey and M. Murray on the role of the
Dixmier-Douady class of a bundle gerbe (this variant of the more general
gerbe theory [Br] was introduced in [Mu]) in quantization; actually,
similar arguments were used previously in [M1] in the study of $3+1$
dimensional Schwinger terms but the role of he DD class was not understood
at that time. The bridge between APS index theory and bundle gerbes
was constructed in [CaMiMu].

\vskip 0.3in
2. RESIDUE CALCULUS OF PSEUDODIFFERENTIAL OPERATORS

\vskip 0.3in
We shall consider differential and pseudodifferential operators
defined in some domain in $\Bbb R^n.$ A PSDO $A$ is defined by giving
its \it symbol \rm $a.$ The symbol is a smooth matrix valued function $a(x,p)$
of coordinates $x=(x_1,\dots, x_n)$ and momenta $p=(p_1,\dots,p_n).$
The action of $A$ onto a square integrable vector valued function in $\Bbb
R^n$ is given by
$$(A\psi)(x) = \frac{1}{(2\pi)^{n/2}} \int a(x,p) e^{-ip\cdot x} \hat\psi(p)dp
\tag2.1$$
where
$$\hat\psi(p)=\frac{1}{(2\pi)^{n/2}}\int e^{ip\cdot x}\psi(x) dx$$
is the Fourier transform of $\psi.$

We shall consider only classical PSDO's; they have an \it asymptotic
expansion\rm, valid for $|p|\mapsto\infty,$
$$a(x,p) \sim \sum_{k=m,m-1,\dots} a_k(x,p)\tag2.2$$
where each $a_k$ is a homogeneous function of degree $k$ in the momenta,
in the sense that $a_k(x,\l p)=\l^k a_k(x,p)$ for any real positive $\l.$
The highest degree $m$ is the order of the PSDO $A$ and $a_m$ is its
\it principal symbol. \rm
Any classical PSDO has a unique asymptotic expansion but two different
PSDO's might have the same asymptotic expansion. The operator is unique
up to addition by a \it infinitely smoothing operator \rm $A.$ The
characteristic property of an infinitely smoothing  symbol is that
it approaches zero faster than inverse of any polynomial as $|p|\mapsto
\infty.$ An example is the operator defined by the smooth symbol $\exp(-
|p|^2).$

Let $A,B$ be a pair of PSDO's with symbols $a,b.$ The product $AB$ has  a
symbol $a*b$ which is given by  the formula
$$(a*b)(x,p)= \sum_m \frac{(-i)^{|m|}}{m!}( \partial^m_p a(x,p))( \partial^m_x
b(x,p)) \tag2.3$$
where $m=(m_1,m_2,\dots,m_m)$ are nonnegative integers, $|m|=m_1+\dots m_n,$
$m!=m_1!m_2!\dots m_n!,$
$\partial^m_p= (\frac{\partial}{\partial p_1})^{m_1}\dots (\frac{\partial}
{\partial p_n})^{m_n},$ and similarly for $\partial^m_x.$

If $A$ is of order $k$ and $B$ is of order $\ell$ then $AB$ is of order
$k+\ell$ and the principal symbol of $AB$ is the matrix product of the
principal symbols of $A,B.$ Note that the subleading term in the commutator
$a_k*b_l-b_l*a_k$ is given by the Poisson bracket $-i\{a_k,b_l\}.$

A PSDO is a trace class operator if its order is strictly less than $-n.$
The trace is then given by     \define\Tr{\text{tr}}
$$\Tr A = \frac{1}{(2\pi)^n} \int \Tr\, a(x,p) dx dp$$
the trace under the integral being the ordinary matrix trace.
There is a usefull alternative linear complex valued functional on the algebra
of PSDO's which also has the characteristic property that the 'trace' of a
commutator is zero. This is the operator residue, due to  V. Guillemin and
M. Wodzicki [Gu],[Wo].
It is simply defined as
$$Res(A) = \frac{1}{(2\pi)^n}\int_{|p|=1} \Tr\, a_{-n}(x,p) dx dS(p)\tag2.4$$
where $dS$ is the volume form on the sphere $|p|=1.$ Alternatively,
$$Res(A)= \frac{1}{(2\pi)^n}\int_S \Tr\, a_{-n}(x,p) dx \theta(d\theta)^{n-1}
\tag2.5$$
where $\theta= \sum p_i dx_i$ is the symplectic 1-form and $S$ is any
surface homotopic to the unit sphere in momentum space. This latter
definition motivates the name operator residue. Let us restate the basic
property
$$Res[A,B]=0\tag2.6$$
for any pair of PSDO's.

The operator residue is closely related to another form of trace: The Dixmier
trace $\Tr_D.$ The latter is defined for operators with a spectrum such that
$$\lim_{\L\to\infty}\int_{|\l|\leq \L} |\l| d\mu(\l)$$
is (at most) logarithmically diverging, where $d\mu$ is the spectral measure.
The Dixmier trace of a positive operator is then defined as the coefficient of
log$\L$ in the above expression as $\L\to\infty.$ The case of a nonpositive
operator needs more care but we shall not discuss that problem here because
we shall need only (2.4).

Suppose now $A$ is a PSDO of order $-n.$ By a simple computation one observes
that each $a_k$ has a finite ordinary trace when $k<-n.$ But the cut-off trace
(momentum cut-off $|p|\leq \L$) of $a_{-n}$ behaves like $log(\L)\times const.$
for large $\L$ where the constant is the integral of $a_{-n}(x,p)$ over the
spherical directions in the momentum space (since $a_{-n}$ is a homogeneous
function of degree $-n$), thus being equal to the residue of $A.$ It follows
that for PSDO's of degree $-n$ the Dixmier trace is the same as the
 residue. For trace class operators both the Dixmier trace and the
residue vanish.

One can also write the operator residue as the complex residue of the
zeta $\zeta_A$ function associated to the operator $A.$ Recall that
$$\zeta_A(s) = \int \l^{-s} d\mu (\l)\tag2.7$$
provided that there is a spectral gap around zero.  This definition is valid
for positive PSDO's of nonzero order and for large  real part of the
complex variable $s.$
By analytic continuation, the definition can be extended to the point $s=0$
and there one can define the zeta function regularized determinant
log(det$_{\zeta} A) = \zeta'_A (0).$ The definition can be even extended
to nonpositive operator provided that there is a cone in the complex plane
not intersecting with the spectrum of $A,$ Friedlander [Fr], Kontsevich,
Vishik [KoVi].

If $A$ is a PSDO of order $k$ then asymptotically $d\mu(\l)\sim |\l|^{(n-k)/k}
d\l$ and therefore the zeta function has a pole at $s=n/k.$ In particular,
there
is a pole at $s=-1$ (corresponding to naive trace of $A$) for $k=-n.$ The
residue
at $s=-1$ is the residue of the operator $A.$

On a vector bundle over an arbitrary compact manifold a PSDO is an operator
which in terms of local coordinates has a representation as (2.1), [H\"o].
Even though the basic formulas of the symbol calculus have only local
meaning, the residue still makes sense and has the cyclic property, [Wo].

\vskip 0.3in
3. THE RADUL COCYCLE

\vskip 0.3in
Let $\gm$ be the Lie algebra of PSDO's on a compact manifold $M.$
We define a central extension $\gm'=\gm\oplus\Bbb C$ of $\gm$ by the
commutators
$$[(A,\l),(B,\mu)] = ([A,B], c(A,B))$$
where $c$ is the Radul cocycle, [Ra] (the one-dimensional case was studied by
Kravchenko and Khesin, [KrKh])
$$c(X,Y)= Res [\text{log}|p|,A]B\tag3.1$$
Here $|p|^2= \sum (p_i)^2$ is the symbol of the euclidean Laplace operator.
Note that log$|p|$ is not a classical PSDO but the commutator $[\text{log}|p|,
A]$ is.

For any PSDO $A$ we have        \redefine\ll{\text{log}|p|}
$$Res[\text{log}|p|,A]=0$$
by integration in parts in configuration space. Therefore
$$0= Res[\text{log}|p|,AB]= Res [\ll,A]B +Res\,A[\ll,B]$$
form which follows
$$Res[\ll,A]B = -Res[\ll,B]A$$
and so $\o(A,B)=-\o(B,A).$ We have to prove the Jacobi identity for $\gm'.$
This is equivalent with the 2-cocycle relation
$$c(A,[B,C])+c(B,[A,C])+c(C,[A,B]).\tag3.2$$
To prove (3.2) we first use the antisymmetry of $\o,$
$$\align c(A,[B,C])&= Res[\ll,A][B,C]= -Res [\ll,[B,C]]A \\
&= -Res[[\ll,B],C] A -Res[B,[\ll,C]]A \\
&= -Res [\ll,B][C,A] -Res [\ll,C][A,B].
\endalign $$
The right-hand side cancels exactly the last two terms in (3.2).

In one space dimension, when using the formula (2.4) for the residue, one
usually modifies the definition of the cocycle:
$$\o(A,B) = \frac12 Res\, \epsilon [\ll,A]B.\tag3.3$$
Here $\epsilon=p/|p|.$ In particular, if $A=A(x)$, $B=B(x)$ are multiplication
operators on the circle,
$$\o(A,B)= \frac{1}{2\pi} \int \Tr A'(x) B(x)\tag3.4$$
is the central term of an affine Kac-Moody algebra.  We shall see that in
applications to quantum field theory in higher space-time dimensions (3.3)
is the correct formula to generalize and not the Radul formula (3.1).

\newpage
4. RENORMALIZATION IN HIGHER DIMENSIONS

\vskip 0.3in

Consider a family of hamiltonians of the form $H(t)=D_0 + A(t)$ acting in
an
one-particle Hilbert space $H.$ We assume that $D_0$ is a time independent
self-adjoint operator and $D_0+A$ is essentially self-adjoint in the
same domain where $A$ is some bounded self-adjoint operator. We shall also
assume that $A(t)$ has compact support in the time variable $t,$ $A(t)=0$ for
$|t| > T.$ We study
the time evolution equation
$$i\partial_t U(t)= H(t)U(t), U(-T)=1 .\tag4.1$$
Writing $V(t)= \exp(it D_0) U(t)\exp(iT D_0)$ we obtain an equivalent equation
$$i\partial_t V(t) = h(t) V(t) \text{ with $V(-T)=1$ }\tag4.2$$
where $h(t)=e^{it D_0} A(t) e^{-it D_0}. $ Since $h(t)$ is bounded and has
finite support in $t$
this equation has a solution for all  times given by the Dyson
expansion
$$ V(t)=1+\sum_{n\geq 1}  (-i)^n \int_{t > t_n \dots  > t_1 >-T} \dots \int
h(t_n) \dots h(t_2)h(t_1) dt_n \dots dt_2 dt_1 \tag4.3$$

Let $\e= D_0/|D_0|.$ We assume that zero is not an eigenvalue of $D_0.$
The spectral decomposition $H=H_+\oplus H_-$ corresponding to the splitting
of the spectrum of $D_0$ to positive and negative parts fixes an irreducible
representation of the canonical anticommutation relations (CAR), uniquely
defined up to equivalence, in a Fock space $\Cal F$ with a vacuum $|0>$
which is annihilated by the elements $a^*(v)$ and $a(u),$ $v\in H_-, u\in
H_+,$ of the CAR algebra
$$a^*(v) a(v') + a(v')a^*(v) = (v,v'),\tag4.4$$
and all the other anticommutators are equal to zero. Let $\{e_n\}_{n\in\Bbb Z}$
be an orthonormal basis of eigenvectors of $D_0$ such that the negative
indices correspond to negative eigenvalues and nonnegative indices correspond
to positive eigenvalues. Set $a_n= a(e_n)$ and $a^*_n = a^*(e_n).$ Fix the
usual normal
ordering for the products of creation and annihilation operators by
$:a^*_n a_m: = -a_m a^*_n$ if $n=m<0$ and all the other products remain
unchanged.

It is known that a bounded one-particle operator $A=(A_{nm})$ can be
canonically quantized as
$$\hat A = \sum A_{nm} : a^*_n a_m :\tag4.5$$
iff $[\e, A]$ is Hilbert-Schmidt, for a survey on these matters see [Ar].
This quantization preserves the
commutation relations of the Lie algebra of linear operators except for
a complex valued cocycle ('Schwinger term') $c_L.$ The Lundberg's cocycle
$c_L$ is given by, [Lu],
$$c_L(X,Y) = \frac14 \Tr\,\epsilon [\e,X][\e,Y]\tag4.6$$
for a pair $X,Y$ of quantizable one-particle operators. The commutation
relations
of the corresponding quantum operators are
$$[\hat X,\hat Y]= \widehat{[X,Y]} +c_L(X,Y).\tag4.7$$

In general, for the interaction term $A,$ $[\e, A]$ is not
Hilbert-Schmidt, $[\e, U]$ is neither Hilbert-Schmidt and therefore the
canonical quantum operator $\hat U$ does not exist.

We assume that the free hamiltonian $D_0$ is a self-adjoint PSDO of
degree $\geq 1.$
We also shall make the following additional assumptions:
\roster \item $A$ is bounded PSDO (thus of degree zero)
\item deg$[|D_0|,A]$ is less or equal to  deg$A.$
\endroster

These properties hold for example when $D_0$ is a Dirac operator and $A=
\g^{\mu} A_{\mu}$ is a gauge interaction term.
Define an unitary operator
$$T_A=\exp(\frac14 |D_0|^{-1}[D_0,A]|D_0|^{-1}).\tag4.8$$
Note that $T_A=1+ K,$ where the degree of $K$ is less or equal to
deg$A-1.$
The operator $T_A$ takes a solution of the time evolution equation
for the hamiltonian $H(t)$ to a solution of the equivalent
hamiltonian $H'(t)= T_A H(t) T_A^{-1} -i(\partial_t T_A)T_A^{-1}.$

\proclaim{Lemma 4.9} With the above assumptions on the interaction $A$ the
operator $[\e,A']$ is a PSDO of degree $k-1,$ where
$A'=H'-D_0$ and the degree of $[\epsilon,A]$ is $k.$ \endproclaim
\demo{Proof} Now
$$A'= T_A A T_A^{-1} - [D_0,T_A]T_A^{-1} -i (\partial_t T_A)T_A^{-1}.$$
Modulo terms of degree $k-1,$
$$[\e,A']= [\e,A]- [D_0,[\e, \frac14 |D_0|^{-1}[D_0,A] |D_0|^{-1} ].$$
On the other hand, the second term is equal to
$$\align \frac14 & [\e, |D_0|^{-1}[D_0,[D_0,A]]|D_0|^{-1}]\\
&= \frac14 [\e, -2\e A\e +\e D_0 A |D_0|^{-1}+|D_0|^{-1}AD_0 \e ]
= \frac14 [\e, -2\e A\e + 2A] + O(k-1)\\
&= [\e,A]+O(k-1)\endalign$$
and thus cancels the $O(k)$  part of the first term. \enddemo

We can apply this method successively: Starting from some interaction $A_0=
A$ such that deg$[\e,A]=k\leq 0$ we get a new interaction $A_1=A'$
using the
conjugation $T_A.$ We can then insert $A_1$ as an argument to $T$ and obtain
an unitary operator $T_{A_1}.$ This defines again a new interaction $A_2$ such
that deg$[\e,A_2]\leq k-2.$ Continuing this way we obtain, after $p$ steps,
an unitary operator $T^{(p)}_A = T_{A_{p-1}}\dots T_{A_0}$ such that the time
evolution for the operator $T^{(p)}_A U(t)$ is determined by an interaction
$A_p$ such that $[\e,A_p]$ is of degree $\leq k-p.$ In particular, for $p>n$
the new interaction satisfies the Hilbert-Schmidt condition. Thus we have,
[LaMi],

\proclaim{Theorem 4.10} There is an unitary PSDO  $T_A,$ depending
continuously on $A,$ such that the time evolution $T_A U(t)$
satisfies the time evolution equation with the hamiltonian $H'(t)=D_0
+A'$ such that $[\e,A']$ is Hilbert-Schmidt. The operator $T_A$ is
obtained as a product of the operators (4.8) obtained recursively from
the initial $A.$ Furthermore, at times $|t| >T$ (for which $A(t)=0$) the
transformation $T_A$ is the identity operator. \endproclaim

This theorem was proven in [LaMi] for writing a simplified (compared with the
original proofs of Palmer and Ruijsenaars [Pa], [Ru]) proof of the existence
of the fermionic scattering matrix in external field problems. The
renormalization which we introduced above has the important property that
the time evolution operator for all intermediate times can be implemented
in the free fermionic Fock space. This is in striking contrast with the
more conventional approaches were only the asymptotic scattering operator
is implementable. We are note going to pursue on the scattering problems
any more here. Instead, we shall use our method in the next section
for a renormalization of the current algebra and calculation of the
Schwinger terms.

\vskip 0.3in
5. SCHWINGER TERMS

\vskip 0.3in
We shall assume that the physical space $M=\Bbb R^n$ and that the vector
potentials, infinitesimal gauge transformations, and the symbols of all other
PSDO's considered below fall off at $|x|\mapsto
\infty$ at least like $|x|^{-n/2 -\a}$ for some positive $\a.$

In this section we shall work in the hamiltonian formalism. At a fixed instant
of time, $t=0$ say, the renormalization operator $T_A$ of the previous section
becomes a time independent function of the potential $A(t=0).$ The hamiltonian
is conjugated as $H'=T_A H T_A^{-1}.$
The currents (multiplication operator in the one-particle space) are also
affected by the renormalization. Before the renormalization the action of a
gauge transformation
$g$ consists of a multiplication operator in the one-particle fermionic Hilbert
space and the usual gauge action in the space of background vector potentials.
After the renormalization in the one-particle space the new fermionic
gauge action is given by the operator
$$ \Omega(g;A)= T_A\, g (T_{A^g})^{-1}.\tag5.1$$
Here $A^g= g^{-1}Ag +g^{-1}dg$ is the gauge transformed potential.
By definition, $\O$ satisfies the one-cocycle equation
$$\O(gg';A)= \O(g;A) \O(g';A^g).\tag5.2$$
Differentiating $\O$ along one-parameter subgroups one obtains the operators
$\theta(X;A)=T_A X T_A^{-1} -(\Cal L_X T_A)T_A^{-1}$ for infinitesimal gauge
transformations $X:M\to \gm.$ Here $\Cal L_X$ is the Lie derivative in the
direction of the infinitesimal gauge transformation $X.$ They satisfy
$$\theta([X,Y];A) = \Cal L_X \theta(Y;A) -\Cal L_Y\theta(X;A)
+[\theta(X;A),\theta(Y;A)].\tag5.3$$
Note that the modified Gauss law generators $G'_X=\Cal L_X +\theta(X;A)$
satisfy
the same commutation relations as the unrenormalized operators $G_X= \Cal L_X+
X.$ This is because actually $G'_X = T_A G_X T_A^{-1}.$

The operators $\theta(X;A)$ satisfy
\proclaim{Lemma 5.4}
$[\e,\theta(X;A)] \in L_2.$  \endproclaim
\demo{Proof} Let $\epsilon(A)$ be the sign of the hamiltonian $D_A=D_0+A;$ we
use the abreviation $A=A_k \g_k.$ Here $\g_k\g_l+\g_l\g_k=2 g_{kl}$ is the
usual $\g$-matrix algebra, $1\leq k,l \leq n.$
Since the commutator of the renormalized interaction $A'$ with
$\e$ is Hilbert-Schmidt, the difference $\e(A)-T_A^{-1}\,\e T_A$
is also Hilbert-Schmidt.

But
$$\align [\e, \O(g;A)]&= \e\, T_A\, g (T_{A^g})^{-1} - T_A\, g (T_{A^g})^{-1}
 \,\e \\
&= T_A\left( ({T_A}^{-1} \,\e\, T_A)g - g((T_{A^g})^{-1} \,\e\, T_A
)\right) {T_A}^{-1}\\
&\equiv T_A \left(\e (A^g) g - g\,\e(A)\right) (T_{A^g})^{-1} \text{ mod }
L_2\endalign$$
where $L_2$ is the ideal of Hilbert-Schmidt operators and we have used the
gauge transformation rule $g^{-1} \e(A)g  =\e(A^g).$
The statement for $\theta(X;A)$ follows by taking the derivative with respect
to
a one-parameter subgroup. \enddemo

{}From (5.4) follows that the canonically quantized operators $\hat
\theta(X;A)$
exist.
The second quantized Gauss law generators $\hat G_X'=\hat\theta(X;A)+\Cal L_X$
act on wave functions $\psi(A)$ taking values in the free fermionic Fock space
$\Cal F.$ The geometric meaning of the renormalization $T$ is this: The
fermionic Fock spaces for each background field $A$ form a vector bundle over
$\Cal A,$ the space of smooth external potentials.
This bundle is trivial since the base is an affine space.
Nevertheless, the representations of the CAR algebra in different fibers
are inequivalent. The renormalization operator $T_A$ identifies the Fock
space over $A$ with the free Fock space over $A=0.$

\bf Remark \rm We could take a more general point of view: Take $\Cal A$ as
the space of all perturbations of the free hamiltonian $D_0$ by bounded
PSDO's satisfying. Assume that the symbol of $D_0$ is $p=p_k \g_k$ plus
operators
of order less or equal to zero, i.e., $D_0$ is a generalized Dirac operator.
The gauge group $\Cal G$ can be taken as the group of unitary PSDO's $g$
such that the principal symbol $g_0$ commutes with $p.$; this
includes in particular the group of ordinary gauge transformations. Lemma
5.4 remains valid in this general setting and the method below for the
computation of the Schwinger terms can be applied.

The generators $\hat G_X'$ satisfy the commutation relations
$$[\hat G_X',\hat G_Y']= \hat G_{[X,Y]}' +c(X,Y;A) \tag 5.5$$
where
$$c(X,Y;A)= c_L(\theta(X;A),\theta(Y;A)). \tag 5.6$$
{}From the definition follows immediately the 2-cocycle property
$$0=c(X,[Y,Z];A) + \Cal L_X c(Y,Z;A) + \text{cyclic permutations}\tag5.7$$

The trace of a PSDO on $\Bbb R^n$ is given as the integral of the total
symbol,
$$\Tr\, A = \frac{1}{(2\pi)^n} \int \Tr\, a(x,p) dx dp \tag5.8$$
where $a$ is the symbol of an operator $A.$ Since the symbol is assumed to be
smooth, the convergence of the integral depends only on the asymptotic
properties
of the symbol. Assuming that the $x$ integral exists, a sufficient condition
for the (absolute) convergence of the trace is that the degree of the
operator is strictly less than $-n.$

It is useful to extend the formula (5.8) to the case when the integral
converges only conditionally in the following sense:
$$\Tr_C \, A= \text{lim}_{\L\to\infty} \int_{|p|<\L} dp \int dx\, \Tr\, a(x,p).
\tag5.9$$

Suppose next that $A$ is a PSDO such that the point-wise trace of the
homogeneous term $a_{-n}$
is a total derivative in momentum space, $\Tr\,a_{-n}= \sum_{k=1}^n
\frac{\partial f_k}{\partial p_k},$
each $f_k$ being homogeneous and of degree $-n+1$ in momenta. Then we can
define
a trace     $\Tr\, A$ in the following way.
For a finite value of the cut-off we have
$$\Tr_{\L}\, A= \sum_{i\leq N} \L^i \a_i(A).\tag5.10$$
In principle we should also have a logarithmically diverging term $log(\L)\cdot
\a_{log}(A)$ arising from the homogeneous symbol of degree $-n.$ However, the
assumption that this particular symbol is a total derivative leads instead to
the finite expression
$$\int_{|p|=\L} \Tr\, \frac{p_k}{|p|} f_k(p,x) dp dx$$
which is independent of the value of the cut-off. It follows that we can define
consistently      \redefine\TR{\text{TR}}
$$\TR \, A = \a_0(A).\tag5.11$$
For trace-class operators $\TR\, A=\Tr\, A$ is the standard trace. Furthermore,
if $A$ is conditionally trace class then $\Tr_C A =\TR\, A.$
(Along these lines there is a  theory of traces of PSDO's of \it nonintegral
order \rm
in a more general setting on a compact manifold by Kontsevich and Vishik,
[KoVi].)

\proclaim{Lemma 5.12} Let $A,B$ a pair of PSDO's on $\Bbb R^n$ (with the
asymptotic $x-$ space properties stated earlier). Then $\TR[A,B]$ exists
and is equal to $Res[log|p|,A] B.$
\endproclaim
\demo{Proof} See [CFNW]. Note that $\TR \, A$ does not exist for all operators
$A$ and therefore the cocycle $\TR[A,B]$ is nontrivial.
\enddemo

\proclaim{Theorem 5.13} Let $\bold b$ be the Lie algebra of all PSDO's of
degree
zero such that $[\e,X]$ is of degree less than $-n/2.$ Then the cocycle $c_L$
is well-defined on $\bold b.$ The function $c(X,Y)= \frac12
Res\, \e[log|p|,X] Y$
is also a 2-cocycle on $\bold b,$ equivalent with $c_L.$ \endproclaim
                                                           
\demo{Proof} In this generality, the theorem was proven in [CFNW];
for the particular case of renormalized gauge currents it was proven
earlier in [M2].  We give here a  simpler
proof.
We can write
$$\align c_L(X,Y)&= \frac12 \Tr_C [\e,X] Y =\frac12 \TR\, [\e,X]Y\\
&=\frac12 \TR\, ( [X\e,Y] -\e[X,Y])
=\frac12 \TR[X\e,Y]-\frac12 \TR\,\e[X,Y].\tag5.14\endalign$$
The first term on the right is in the TR class because it is a commutator.
Since we know that the sum is conditionally covergent, the second term is
also in the TR class. We can define an extension $\TR'$ of TR to the space of
all PSDO's. This is of course not unique, but any extension will do.
Define $\l(X)= \TR ' \e X.$ Obviously $\TR \,\e[X,Y]$ is a coboundary of
$\l.$ Thus the Lundberg cocycle is cohomologous to the first term on the
right in (5.14). Using (5.12) we obtain
$$c_L(X,Y) \sim -\frac12 Res [log|p|,Y]X\e=-\frac12 Res\,\e [log|p|,Y]X
= \frac12 Res\,\e[log|p|,X]Y.  $$
\enddemo

Now we can apply this general formula to the case of renormalized current
operators. For example, in three space dimensions and in the case of
2-component
fermions we can choose $T_A$ such
that it has the following asymptotic expansion:
$$T_A = 1+ \frac14 \frac{[p,A]}{|p|^2} +O(-2). \tag5.15$$
Here $p=p_k\sigma_k$ and $A=A_k\sigma_k$ and the $\sigma_k$'s are the hermitian
Pauli matrices, $k=1,2,3.$
We have dropped the lower order terms because they are already Hilbert-Schmidt.
Of course, they must be included in order to preserve unitarity. For
calculating
Schwinger terms we do not need any terms which are of order less than $-2.$
This is because the residue in theorem (5.13) depends only on the degree
three term; on the other hand, the commutator with $log|p|$ introduces
one extra factor $1/|p|,$ so $X,Y$ need to be known only down to degree
$-2.$

The renormalized currents are given then as ($X,Y$ are now multiplication
operators by Lie algebra valued functions)
$$\theta(X;A)=  X + \frac{i}{4|p|^2} [p,dX] -\frac{1}{4|p|^2}[\sigma_k,A]
\partial_k X +\frac{[p,A]}{2|p|^4} p_k\partial_k X
+\frac{[p,A]}{16|p|^4}[p,dX] + O(-3) .\tag5.16$$
Here $dX=\sigma_k\partial_k X.$
Inserting $\theta(X;A)$ and $\theta(Y;A)$ as the arguments in the cocycle,
instead of $X,Y,$ one obtains after a straight-forward computation, [Mi2],
$$c(X,Y;A)= \frac{i}{24\pi^2} \int \Tr A\wedge(dX\wedge dY- dY\wedge dX).
\tag5.17$$
This is the Faddeev-Mickelsson cocycle found in another context in [Mi3],
[F-Sh].

\vskip 0.3in

6. GERBES, INDEX THEORY, ANOMALIES

\vskip 0.3in
In this section we want to describe a more geometric approach to the
problem of construction of the family of Fock spaces parametrized by
external vector potentials and the action of the gauge group. Actually,
the method here is very general and applies as well to the case of an
external metric field or any other interactions for that matter. In order to
keep the discussion as simple as possible we shall restrict to the case
of vector potentials.

The fermionic Fock spaces parametrized by Yang-Mills potentials form a
vector bundle $\Cal F$ over the space $\Cal A.$ In the case of chiral
massless fermions there are subtleties in defining this bundle.
The difficulty is related to the fact that the splitting of the one particle
fermionic Hilbert space $H$ to positive and negative energies is not a
continuous function of the external field.
One can easily construct paths in the space of external fields
such that at some point on the path a positive energy  state dives into
the negative energy space (or vice versa). These points are obviously
discontinuities in the definition of the space of negative energy states
and therefore the fermionic vacua do not form of smooth vector bundle over the
space of external fields. This problem does not arise if we have massive
fermions in the temporal gauge $A_0=0.$ In that case there is a mass gap
$[-m,m]$ in the spectrum of the Dirac hamiltonians and the polarization
to positive and negative energy subspaces is indeed continuous.

If $\l$ is a real number not in the spectrum of the hamiltonian then one
can define a bundle of fermionic Fock spaces $\Cal F'_{A,\l}$ over the set
$U_{\l}$ of external fields $A,$ $\l\notin Spec(D_A).$ The vacuum in
$\Cal F'_{A,\l}$ is defined by the polarization of the one-particle space
to positive and negative spectrum of the operator $D_A-\l.$ It turns out that
the
Fock spaces $\Cal F'_{A,\l}$ and $\Cal F'_{A,\l'}$ are naturally isomorphic
up to a phase. The phase is related to the arbitrariness in filling the
Dirac sea between vacuum levels $\l,\l'.$ Such a filling is given corresponds
(because of the anticommutation relations) to an exterior product
$v_1\wedge v_2\wedge\dots v_m$ of a complete
orthonormal set of eigenvectors $D_A v_i= \l_i v_i$ with $\l<\l_i <\l'.$
A rotation of the eigenvector basis gives a multiplication of the exterior
product by the determinant of the rotation. Thus there is a well-defined
complex line $DET_{\l\l'}(A)$ for each $A\in U_{\l}\cap U_{\l'}=U_{\l\l'}$ and
$$\Cal F'_{A,\l'}= \Cal F'_{A,\l} \otimes DET_{\l\l'}(A)\tag6.1$$
over the intersection set. We set $DET_{\l'\l}=DET_{\l\l'}^{-1}$ for
$\l <\l'.$ Note that from these definitions follows immediately
that the line $DET_{\l\l''}$ can be naturally identified as $DET_{\l\l'}\otimes
DET_{\l'\l''},$ i.e., the local line bundles $DET_{\l\l'}$ form a cocycle
over the open cover $\{U_{\l}\}$ of $\Cal A.$
In order to compensate the dependence on $\l$ in the definition of the
Fock spaces we search for a family of complex line bundles $DET_{\l}$
over the open sets $U_{\l}$ such that
$$DET_{\l'}= DET_{\l'\l} \otimes DET_{\l}\tag6.2$$
over $U_{\l\l'}.$ Obviously, the cocycle property of the of the line bundles
$DET_{\l\l'}$ is a necessary condition for the existence of the family of
bundles $DET_{\l}.$ It is not very hard to prove that this is also a sufficient
condition, [Mi1]. This follows also form the general theory of bundle gerbes
[Mu] since $\Cal A$ is topologically trivial.

We  define  the tensor product
$$\Cal F_{A,\l}=\Cal F'_{A,\l} \otimes DET_{A,\l}.\tag6.3$$
Using (6.1) and (6.2) we observe that the right-hand side is independent
from $\l$ and one has a well-defined bundle
$\Cal F$ of Fock spaces over all of $\Cal A.$

Next one can ask what is the action of the gauge group in $\Cal F.$
The gauge action in $U_{\l}$ lifts naturally to $\Cal F'.$ Thus the only
problem is to construct a lift of the action on the base to the total
space of $DET_{\l}.$ Note that the determinant bundle here is a bundle over
external fields in \it odd dimension, \rm and therefore one would expect that
it is trivial (curvature equal to zero) on the basis of families index
theorem. However, it turns out that the relevant determinant bundle
actually comes from a determinant bundle in even dimensions. Instead
of single vector potentials we must study paths in $\Cal A,$ thus the
extra dimension. The relevant index theorem is then the APS theorem
for even dimensional manifolds with a boundary; physically, the boundary
can be interpreted as the union of the space at the present time and
in the infinite past, [CaMiMu].

We recall some facts about lifting a group action on the base space $X$ of a
complex line bundle to the total space $E.$ Let $\omega$ be the curvature
2-form of the line bundle. It is integral in the sense that $\int \omega$
over any cycle is $2\pi \times$ an integer. Let $G$ be a group acting
smoothly on $X.$ Then there is an extension $\hat G$ which acts on $E$ and
covers the $G$ action on $X.$ The fiber of $\hat G \to G$ is equal to
$Map(X,S^1).$ As a vector space, the Lie algebra of the extension is
$\gm \oplus Map(X,i\Bbb R).$ The commutators are defined as
$$[(a,\a),(b,\b)]= ([a,b],\o(a,b)+\Cal L_a \b-\Cal L_b \a)\tag6.4$$
where $a,b\in \gm$ and $\a,\b: X\to i\Bbb R.$ The vector fields generated
by the $G$ action on $X$ are denoted by the same symbols as the Lie algebra
elements $a,b;$ thus $\o(a,b)$ is the function on $X$ obtained by evaluating
the 2-form $\o$ along the vector fields $a,b.$ The Jacobi identity
$$\o([a,b],c) + \Cal L_a \o(b,c) + \text{cyclic permutations} =0$$
for the
Lie algebra extension $\hat\gm$ follows from $d\o=0.$

What we need is a formula for the curvature of the line bundles $DET_{\l}$
along gauge directions. Not surprisingly, this is given by a reduction from
a secondary characteristic class. Recall that in even space-time dimensions
the Chern class of the determinant line bundle is obtained by starting from
an appropriate characteristic class (the class appearing in the index formula
of Dirac operators) in two higher dimensions and then integrating
over the space-time manifold; this leaves a closed integral differential form
of degree two on the parameter space of the Dirac operators, [AS]. In the odd
dimensional case here one starts from the APS index formula on a manifold
with a boundary, [APS]. The formula contains two pieces on the right-hand side.
The first is an integral of a local differential polynomial (the same as in the
case without boundary) and the second is the so-called eta-invariant which
contains nonlocal information about the spectrum of the boundary Dirac
operator. The essential property of the eta-invariant is that it is
gauge invariant. For that reason it does not give a contribution to the
curvature of the determinant bundle along gauge orbits. Everything comes from
the local differential polynomial; the non- gauge invariant piece of the latter
comes from the boundary and is equal to a secondary characteristic class. In
simple situations this is just a Chern-Simons form.

Integrating the Chern-Simons form in $2n+3$ dimensions over the $2n+1$
dimensional
physical space gives a 2-form along gauge orbits.

For example, when dim$M=1$, starting from the Chern-Simons form $\frac{1}{8\pi^
2}\Tr(AdA + \frac23 A^3)$ we get
$$\omega_A(X,Y) = \frac{1}{4\pi} \int_{S^1} \text{tr}\, A_{\phi}
[X,Y],\tag6.5$$
the curvature at the point $A$ in the directions
of infinitesimal gauge transformations $X,Y.$ (Note the normalization factor
$2\pi$ relating the Chern class to the curvature formula.)
This is not quite the
central term of an affine Kac-Moody algebra, but it is equivalent to it
(in the cohomology with coefficients in Map$(\Cal A,\Bbb C)).$ In other
words, there is a 1-form $\theta$ along gauge orbits in $\Cal A$ such that
$d\theta= \o-c ,$ where
$$c(X,Y)= \frac{i}{2\pi} \int \Tr X \partial_{\phi} Y\tag6.6$$
is the central term of the Kac-Moody algebra, considered as a closed
constant coefficient 2-form on the gauge orbits. There is a simple explicit
expression for $\theta,$
$$\theta_A (X)= \frac{i}{4\pi} \int \Tr A X.$$

When dim$M=3$ the curvature (or equivalently, the Schwinger term) is
obtained from the five dimensional Chern-Simons form
$$CS_5(A)= \frac{i}{24\pi^3}\Tr (A (dA)^2  + \frac32 A^3 dA + \frac35 A^5).$$
By the same procedure as in the one dimensional case we obtain
$$\omega_A(X,Y)= \frac{i}{4\pi^2}\int \Tr \left((AdA+dA\,A+A^3)[X,Y]+XdA\,YA
-YdA\,XA \right)  . \tag6.7$$
This differs from the FM cocycle
$$\omega'_A(X,Y)=\frac{i}{24\pi^2} \int \Tr A (dX dY - dY dX)$$
by the coboundary of
$$\frac{-i}{24\pi^2}\int \Tr (AdA +dA\,A+A^3) X .$$

\vskip 0.3in
\it The Dixmier-Douady class \rm

\vskip 0.3in
Let $P\Cal F$ be the bundle of \it projective \rm Fock spaces $\Cal F_A/\Bbb
C^{\times}$
over $\Cal A.$ Because the action of $\Cal G$ on $\Cal A$ lifts to the total
space $\Cal F$ modulo $A$-dependent phases arising from the Schwinger terms
the group $\Cal G$ acts on $P\Cal F.$ The action of the subgroup $\Cal G_0$
of based gauge transformations is free and therefore we can define the
quotient bundle $P\Cal F/\Cal G_0$ over the manifold $\Cal A/\Cal G_0.$
This projective bundle is nontrivial in the sense that there is no Hilbert
bundle $\Cal H$ over $\Cal A/\Cal G_0$ whose projectivization would be
equal to $P\Cal F/\Cal G_0,$ [Se]. The obstruction to constructing the
Hilbert bundle is a certain element $\o_3\in H^3(\Cal A/\Cal G_0,\Bbb Z),$
called the Dixmier-Douady class of the bundle, [Br]. The relation of
the DD class to quantum field theory was recently clarified by Carey and
Murray, [CaMu]; see also [CaMuWa].
I shall briefly describe the construction of $\o_3$ below.

We need a locally finite partition of unity on $X=\Cal A/\Cal G_0$ subordinate
to the
covering by the open sets $V_{\l}=\pi(U_{\l})$ where $\pi:\Cal A\to \Cal A/
\Cal G_0$ is the projection. On a locally compact manifold there exists
always a partition of unity subordinate to a given open cover. However, in this
case $X$ is not locally compact and we have no proof of the existence of the
partition of unity. For that reason we assume that $X$ stands for any finite-
dimensional submanifold of $\Cal A/\Cal G_0$ or any other submanifold such
that there is a partition of unity $\{f_{\l}\}$ subordinate to the open sets
$X\cap V_{\l}.$ Let $\theta_{\l\l'}$ be a representative for the Chern class
of the bundle $DET_{\l\l'}.$ Because of the cocycle property of the line
bundles we can choose the 2-forms $\theta_{\l\l'}$ such that
$$\theta_{\l\l'} +\theta_{\l'\l''}= \theta_{\l\l''}\tag6.8 $$
The forms $\theta_{\l\l'}$ on $U_{\l\l'}$ descend to forms on $V_{\l\l'}.$

We define
$$\theta_{\l}(x) = \sum_{\l'} \theta_{\l\l'}(x) f_{\l'}(x)\tag6.9$$
at points $x\in V_{\l}.$  Now we have
$$\theta_{\l} -\theta_{\l'} = \sum_{\l''} (\theta_{\l\l''}-\theta_{\l'\l''})
f_{\l''}= \sum_{\l''} \theta_{\l\l'} f_{\l''}= \theta_{\l\l'}\tag6.10$$
by (6.8) and by $\sum f_{\l}(x)=1,$ on the intersection $V_{\l\l'}.$
The forms $\theta_{\l}$ are not closed but on the intersection $V_{\l\l'}$
we have $d\theta_{\l}= d\theta_{\l'}$ since $\theta_{\l\l'}$ is closed.
Thus we may patch together the closed forms $d\theta_{\l}$ to a global
closed form $\o_3$ on $X.$ This is the Dixmier-Douady class of the
projective bundle $P\Cal F.$

The above construction is an example of a \it bundle gerbe, \rm introduced
by Murray, [Mu] (which in turn is a specialization of the more general
theory of gerbes, [Br]). A bundle gerbe is defined as follows. Let
$\pi: Y \to X$ be some fibration; in general this is not locally trivial, so
$Y$ does not need to be a fiber bundle over $X.$ Let
$$Y^{[2]}= \{(y,y')\in Y\times Y| \pi(y)=\pi(y')\}.$$
A bundle gerbe is a principal $\Bbb C^{\times}$ bundle $P$ over $Y^{[2]}$ with
a smooth associative composition map
$$P_{(x,y)}\times P_{(y,z)}\to P_{(x,z)}.$$
The bundle gerbe has also an identity (which is a section of $P$ over the
diagonal $Y\subset Y^{[2]}$) and an inverse $P_{(x,y)} \to P_{(y,x)},$
$p\mapsto p^{-1}.$

\bf Example \rm Let $X=\Cal A$ and $Y=\{(A,\l)| A\in\Cal A,\,\, \l\notin
Spec(D_A)\}.$ $\pi:Y\to X$ is the natural projection. In this case $Y^{[2]}=
\{(A,\l,\l')| \l,\l'\notin Spec(D_A)\}.$ The bundle $P$ over $Y^{[2]}$ is
obtained as the collection of the line bundles $DET_{\l\l'}$ (with the zero
section deleted) over the sets
$\{(A,\l,\l')| A\in U_{\l\l'}\}\subset Y^{[2]}.$
Similarly, a curvature form $\theta$ for $P$ is obtained by patching together
the local 2-forms $\theta_{\l\l'}.$
The fiber product is given
by the natural identification of $DET_{\l\l'} \otimes DET_{\l'\l''}$ and
$DET_{\l\l''}.$  Since the bundles $DET_{\l\l'}$ descend to $V_{\l\l'}$ and
the forms $\theta_{\l\l'}$ are gauge invariant, the whole construction descends
to the quationt by $\Cal G_0$ producing a bundle gerbe over $\Cal A/\Cal G_0.$

In the above example the bundle gerbe over $\Cal A$ is trivial (since $\Cal A$)
is flat, which means that
$$P= {\pi_1}^*(L) \otimes {\pi_2}^*(L^{-1})$$
for some line bundle $L$ over $Y.$ In our example $L$ is obtained by patching
together the local bundles $DET_{\l}.$ However, the corresponding bundle gerbe
over $\Cal A/\Cal G_0$ is nontrivial. The obstruction to trivializing $P/\Cal
G_0$ is given by the Dixmier-Douady class $[\o_3].$

In general, the DD class is constructed starting from the short exact
sequence  of de Rham complexes
$$ 0 \rightarrow \O^*(X) \overset{\pi^*}\to{\rightarrow} \O^*(Y) \overset{
{\pi_1}^* -{\pi_2}^*}\to{\rightarrow} \O^*(Y^{[2]})\cap Im({\pi_1}^* -
{\pi_2}^*) \rightarrow 0.$$
This induces a long exact sequence in cohomology:
$$\dots\rightarrow H^q(X) \overset{\pi^*}\to{\rightarrow} H^q(Y)\overset{
{\pi_1}^* -{\pi_2}^*}\to{\rightarrow} {H^q}_{\pi}(Y^{[2]}) \overset{\Delta}
\to{\rightarrow} H^{q+1}(X)\rightarrow\dots$$
where  $H^q_{\pi}$ denotes the image of ${\pi_1}^* -{\pi_2}^*.$ The form
$\o_3$ which we constructed above is actually equal to $\Delta(\theta).$
However, the same reservation applies to the use of the exact sequence as
we had before in the construction of $\o_3:$ The construction of the map
$\Delta$ uses a locally finite open cover, so strictly speaking it is
valid only in the case of locally compact manifolds.

Finally, I want to mention an important topic closely related to the
problems discussed in these notes but which were omitted here because of
lack of time and space. The structure of the current algebra and anomalies
in gauge
theory can be described in a more general setting of Fredholm modules,
incorporating ideas from noncommutative geometry, [Co]. This was initiated
in [MiRa] (see also the monograph in [Mi3]) and generalized even further
in recent articles, see [La] and references therein.

\vskip 0.3in
\bf Acknowledgements \rm I wish to thank A. Carey, D. Burghelea,
E. Langmann, and S.N.M. Ruijsenaars for discussions at various stages in
preparing these
lecture notes. I also want to thank G. Marmo and P. Michor for inviting
me to the Erwin Schr\"odinger Institute, were the main part of this work
was written.

\vskip 0.3in
REFERENCES

\vskip 0.3in

[APS] M. Atiyah, Patodi, and I. Singer: Spectral asymmetry and Riemannian
geometry I-III. Math. Proc. Camb. Phil. Soc. \bf 77, \rm 43 (1975);
\bf 78, \rm 405 (1975); \bf 79, \rm 71 (1976)

[Ar] H. Araki: Bogoliubov automorphisms and Fock representations of
canonical anticommutation relations   In: Contemporary Mathematics, vol.
\bf 62, \rm American Mathematical Society, Providence (1987)

[AS] M. Atiyah, I. Singer: Dirac operators coupled to vector potentials.
Proc. Natl. Acad. Sci. USA, \bf 81, \rm 2597 (1984)
[Br] J.-L. Brylinski: \it Loop Spaces, Characteristic Classes, and
Geometric Quantization. \rm Birkh\"auser, Boston-Basel-Berlin (1993).

[CaMiMu] A.L. Carey, J. Mickelsson, M. Murray: Index theory, gerbes, and
quantization. In preparation.

[CaMu]  A.L. Carey and M.K. Murray.: Mathematical remarks on the cohomology of
gauge groups and anomalies. To appear in Int J. Mod. Phys. A. hep-th/9408141.

[CaMu1]  A.L. Carey and M.K. Murray.: Faddeev's anomaly and bundle gerbes.
To appear in Lett. Math. Phys.

[CaMuWa] A.L. Carey, M.K. Murray and B. Wang.: Higher bundle gerbes, descent
equations and 3-Cocycles, preprint 1995.

[CFNW] M. Cederwall, G. Ferretti, B. Nilsson, and A. Westerberg: Schwinger
terms and cohomology of pseudodifferential operators.
hep-th/9410016

[Co] A. Connes: \it Noncommutative Geometry. \rm Academic Press (1994)

[F-Sh] L. Faddeev:  Operator anomaly for the Gauss law.  Phys. Lett. \bf B 145,
\rm  81 (1984). L. Faddeev and
S. Shatasvili Theoret. Math. Phys. \bf 60, \rm 770 (1984)

[Fr] L. Friedlander: PhD thesis, Dept. of Math., M.I.T. (1989)

[Gu] V. Guillemin: A new proof of Weyl's formula on the asymptotic distribution
of eigenvalues. Adv. Math. \bf 55, \rm 131 (1985)

[H\"o] H\"ormander:  \it The Analysis of Partial Differential Operators III.
\rm Springer-Verlag, Berlin (1985)

[KoVi] M. Kontsevich, S. Vishik:  Determinants of elliptic pseudo-differential
operators. hep-th/9404046

[KrKh] O.S. Kravchenko and  B.A. Khesin: A nontrivial central extension of
the Lie algebra of pseudodifferential symbols on the circle. Funct. Anal.
Appl. \bf 25, \rm 83 (1991)

[La] E. Langmann: Noncommutative integration calculus. J. Math. Phys. \bf 36,
\rm 3822 (1995). Descent equations of Yang-Mills anomalies in noncommutative
geometry. hep-th/9508003

[LaMi]  E. Langmann and J.  Mickelsson: Scattering matrix in external fields.
hep-th/9509?

[Lu] L.-E. Lundberg: Quasi-free second quantization. Commun. Math. Phys.
\bf 50, \rm 103 (1976)

[Mi1] J. Mickelsson: On the hamiltonian approach to commutator anomalies
in $3+1$ dimensions. Phys. Lett. \bf B 241, \rm 70  (1990)

[Mi2] J. Mickelsson: Wodzicki residue and anomalies of current algebras.
ed. by A. Alekseev et al. Springer Lecture
Notes in Physics 436 (1994)

[Mi3] J. Mickelsson:  Chiral anomalies in even and odd dimensions.
Commun. Math. Phys. \bf 97, \rm 361 (1985). \it Current Algebras and Groups.
\rm
Plenum Press, London and New York (1989)

[MiRa] J. Mickelsson and S. Rajeev: Current algebras in $d+1$ dimensions and
determinant bundles over infinite-dimensional Grassmannians. Commun. Math.
Phys. \bf 116, \rm  365 (1988)

[Mu]  M.K. Murray.: Bundle gerbes. dg-ga/9407015. To appear in  Journal of
the London Mathematical Society.

[Pa] J. Palmer: Scattering automorphisms of the Dirac field. J. Math. Anal.
Appl. \bf 64, \rm 189 (1978)

[Ra] A.O. Radul: Lie algebras of differential operators, their central
extensions, and W-algebras. Funct. Anal. Appl. \bf 25, \rm 33 (1991)

[Ru] S.N.M. Ruijsenaars: Gauge dependence and implementability of the
$S$-operator for spin-0 and spin-$\frac12$ particles in time-dependent
external fields. J. Funct. Anal. \bf 33, \rm 47 (1979)

[Se] G. Segal, unpublished preprint, Dept. of Math., Oxford University (1985)

[Wo] M. Wodzicki: Noncommutative residue. In: Springer Lecture Notes
in Mathematics 1289, ed. by Yu.I. Manin (1984)

\enddocument